# Beating bandwidth limits for large aperture broadband nano-optics

*Johannes E. Fröch[1,2,*, ɫ], Praneeth K. Chakravarthula[3, ɫ], Jipeng Sun[3], Ethan Tseng[3], Shane Colburn[2,4], Alan Zhan[4], Forrest Miller[2], Anna Wirth-Singh[1], Quentin A.A. Tanguy[2], Zheyi Han[2], Karl F. Böhringer[2,5,6], Felix Heide[3], Arka Majumdar[1,2,*]*

1: Department of Physics, University of Washington, Seattle, 98195, WA, USA

2: Department of Electrical and Computer Engineering, University of Washington, Seattle, 98195, WA, USA

3: Department of Computer Science, Princeton University, Princeton, 08540, NJ, USA

4: Tunoptix, 4000 Mason Road 300, Fluke Hall, Seattle, WA, 98195 USA

5: Department of Bioengineering, University of Washington, Seattle, WA, 98195, USA

6: Institute for Nano-Engineered Systems, University of Washington, Seattle, WA, 98195, USA

ɫ : equal contribution

email: jfroech@uw.edu, arka@uw.edu

**Abstract**

Flat optics have been proposed as an attractive approach for the implementation of new imaging and sensing modalities to replace and augment refractive optics. However, chromatic aberrations impose fundamental limitations on diffractive flat optics. As such, true broadband high-quality imaging has thus far been out of reach for low f-number, large aperture, flat optics. In this work, we overcome these intrinsic fundamental limitations, achieving broadband imaging in the visible wavelength range with a flat meta-optic, co-designed with computational reconstruction. We derive the necessary conditions for a broadband, 1 cm aperture, f/2 flat optic, with a diagonal field of view of 30° and an average system MTF contrast of 30% or larger for a spatial frequency of 100 lp/mm in the visible band (> 50 % for 70 lp/mm and below). Finally, we use a coaxial, dual-aperture system to train the broadband imaging meta-optic with a learned reconstruction method operating on pair-wise captured imaging data.

Fundamentally, our work challenges the entrenched belief of the inability of capturing high-quality, full-color images using a single large aperture meta-optic.

## Introduction

Cameras have become omnipresent in contemporary lives, whether in laptops, smartphones, automotive sensors, medical instruments, or surveillance.(*1–3*) While imaging methods are proposed to achieve increasingly higher image quality or extract more information, the miniaturization of optical elements has become equally, if not a more important challenge for emerging applications. For instance, smaller lens systems could facilitate less invasive surgeries with optical endoscopes, lightweight optical systems could reduce the power consumption of airborne drones or satellites, and even the lens bump in mobile phone cameras, ultimately limiting the thickness of smartphones could be eliminated. To this end, flat diffractive optical elements promise to replace or augment refractive optics.

Specifically, meta-optics - sub-wavelength, diffractive, quasi-periodic arrays of nano-scale scatterers, can alter the transmitted wavefront by locally imparting a designed phase delay onto the light field. Thus, by global engineering of the scatterer arrangement, a desired phase profile can be implemented, enabling the deliberate manipulation of light according to a specific optical function. Given this potential of meta-optics,(*4–6*) an emerging body of work has explored their capabilities, rendering exotic functionalities, such as for holography,(*7–9*) multi-modal sensing applications,(*10–16*) spectrometers and hyperspectral imagers, (*17–20*) or optical encoders for hybrid neural networks. (*21–24*)

Diffractive optics, however, have fundamental limitations. To reduce the thickness, the absolute phase is wrapped to a range of (typically) 0 – 2π. This phase wrapping inherently results in chromatic aberrations that are significantly larger than those of standard refractive lenses, outweighing the chromatic response of the scatterer.(*25*, *26*) Thus, one can design a meta-optic that produces a diffraction-limited point spread function (PSF) at a specific wavelength but will perform poorly over an extended bandwidth at the same focal length. For large apertures, it is fundamentally not possible to obtain a high Strehl ratio (> 0.8) of the standalone meta-optic for a large spectral bandwidth and useful numerical aperture (NA), as detailed in several works.(*27*, *28*) This has limited current broadband meta-optics to either small apertures (~100 µm), low NAs (~ 0.05), or strongly limited resolution. (*29*–*33*)

To overcome this fundamental physical limitation of meta-optics, several efforts have investigated high-quality imaging by computational reconstruction or arrays of meta-optics.(*34*–*39*) Images captured by the meta-optic are passed to a computational back end, and the final image quality is improved through aberration correction. Yet, no work with or without a computational back end has demonstrated broadband, full-color imaging with a single meta-optic at a large aperture (> 4 mm), suitable for integration with consumer electronics. Existing works that utilized computational post-processing also lacked an understanding of an intuitive appropriate design approach. We find that <u>understanding why certain designs of meta-optics provide enhanced image quality and why these are more amenable for computational post-processing</u> enables general functional meta-optics.

Specifically, we demonstrate polarization-insensitive, full-color, broadband imaging with a single large aperture (1 cm diameter, f/2 lens) meta-optic. To optimize this large meta-optic, we gradually evolve the design complexity in a constrained parameter space from a hyperboloid phase towards an optimized end-to-end broadband design. This enables image capture in the broadband visible range (~ 400 nm – 700 nm) with video rates at a full diagonal field of view of ~$30^o$. Although the standalone meta-optic is not performing at the diffraction limit, the co-designed computational back-end augments aberration correction, and thus facilitates high-quality imaging, on par with a single refractive lens of the same diameter and f-number. Even more strikingly, for large field angles and depth of field, the meta-optic outperforms the single refractive optic. We further leverage recent breakthroughs in computer science to reinforce the computational backend utilizing a probabilistic diffusion-based reconstruction method. As our meta-optic can be directly integrated with a camera, we implemented a dual-aperture configuration to train and verify the learned backend. Through this approach, we ultimately demonstrate imaging quality comparable to that of current generation compound smartphone lenses. In essence, we report the first large aperture meta-optic with a diameter of 1 cm and f-number of 2, suitable for direct camera integration, with short exposure times and applicable to broad band full color imaging under ambient light and arbitrary scenes.

**Results**

While the design of refractive optics and optical assemblies are well developed, thanks to sophisticated ray tracing software, optimization of complex meta-optics ideally requires a form of full-wave simulation to accurately capture their diffractive nature and to account for the sub-wavelength scatterers. However, this amounts to solving a computationally prohibitive problem: a single meta-optic with a scatterer spacing of ~ 250 nm, and 1 cm circular aperture requires the computation of ~ $10^9$ scatterers (Figure 1a), equal to ~ 7 GB of RAM to analyze their performance using the angular spectrum method, which can account for propagation of such a phase mask. This number linearly increases with the number of wavelengths and field angles that are sampled for the optimization. Although a single simulation for such a device is possible, optimization typically involves tens of thousands of simulations to achieve sufficient quality.

We assume a radially symmetric phase profile and propagation conditions, reducing the design parameters to $\sim 10^4$. Although this restricts the optimization to only on-axis field propagation, we later demonstrate that this is a fair restriction, because emerging designs ensure both spectral broadband, and reasonable field angle performance. We optimize the meta-optics in two consecutive steps. First, as outlined in Figure 1b, we use a differentiable pipeline, where the phase profile of the meta-optic is iteratively updated to maximize the intensity of the point spread function across a spectral range of 450 nm – 650 nm, with a sampling rate of 0.1 nm (details in the Method). This technique exploits symmetry of the meta-optic as well as memory reduction associated with only needing to compute the intensity at focus to enable efficient optimization of a 1 cm aperture with a dense sampling of 2000 wavelengths across the visible band, utilizing a TensorFlow implementation of the Rayleigh-Sommerfeld diffraction integral to optimize the meta-optic using automatic differentiation. This initial optimization step yields a meta-optic with an extended depth of field (EDOF) profile, featuring a focus profile that is elongated over a certain distance along the optical axis, as shown for different wavelengths in Figure 1b. Intuitively, we can understand the broadband functionality of this design emerging, due to the inverse dependence of focal length to the wavelength, akin to the focal shift of a standard hyperboloid metalens.(25, 26) However, in contrast to a hyperboloid metalens, as the focus profile extends over a longer depth, a similar point spread function (PSF) is maintained at the image plane over a broader spectral range.

While this initial optimization step generates a design suitable for broadband imaging (as shown later), we further optimize it in an end-to-end optimization approach, to co-optimize the meta-optic with a complementary computational backend. As illustrated in Figure 1b, the meta-optic is co-designed with a physics-informed computational backend with a learned probabilistic prior-based reconstruction method (details in the Method). This step ensures that the optimized meta-optic exhibits a PSF that is amenable for computational reconstruction.

End-to-end design was performed for two specific applications; one to achieve high-quality imaging for 3 distinct wavelengths in the red, green, and blue range (i.e. a polychromatic design); the second one for broadband operation, with a sampling of ~ 5 nm over the whole visible range. While the former could for instance be implemented in applications where the spectrum of the light source can be controlled (e.g., virtual reality visors), the latter is suitable for broadband imaging under ambient light conditions. Further results and discussion on the optimization are provided in the Supplementary Information. An image of the fabricated broadband meta-optic, held towards a vivid test scene is shown in Figure 1c, illustrating their lensing capability, while all colors are preserved across the visible spectrum. All presented meta-optics considered in this work were fabricated in a SiN-on-quartz platform, fully detailed in the methods section and Supplementary Information.

A side-by-side comparison with a refractive lens of the same aperture and f-number (Figure 1d) highlights the size reduction achieved by the meta-optic. We note that the thickness of the meta-optic is on the order of 1 μm, thus demonstrating a thickness reduction of 4 orders of magnitude (considering the sagitta of the refractive lens). Importantly, we demonstrate that despite orders of magnitude reduction in size, parity in image quality is possible and can even outperform the refractive lens in some cases. The uniform structural quality of the meta-surface after fabrication is apparent as a radially equal structural color effect, seen in optical microscope images (Figure 1e), and by direct inspection of the nanostructure in a Scanning Electron Microscope from top and oblique angles as shown in Figures 1f, and 1g, respectively. Further details and characterization are presented in the Supplementary Information.

We integrated the flat nano-optics directly with a commercial camera (Allied Vision, ProSilica GT1930) using a 3D-printed aperture, shown in Figure 1h. Without the aid of any relay optic, we can directly assess the imaging performance from a system-level perspective. Previously employed relay configurations, as often used in meta-optic works,(*30*, *31*, *34*, *37*) required long exposure times due to their reduced NA relative to the meta-optic and the effective reduction in pixel size with the relay's associated magnification. The large aperture size of the meta-optic in this work instead enables imaging at high framerates as the entire sensor is utilized without an NA-mismatched objective or extra magnifying optics in the path. We further emphasize that by benchmarking against an equivalent refractive lens, we make a fair head-to-head comparison between meta-optics and a refractive lens.(*40*, *41*)

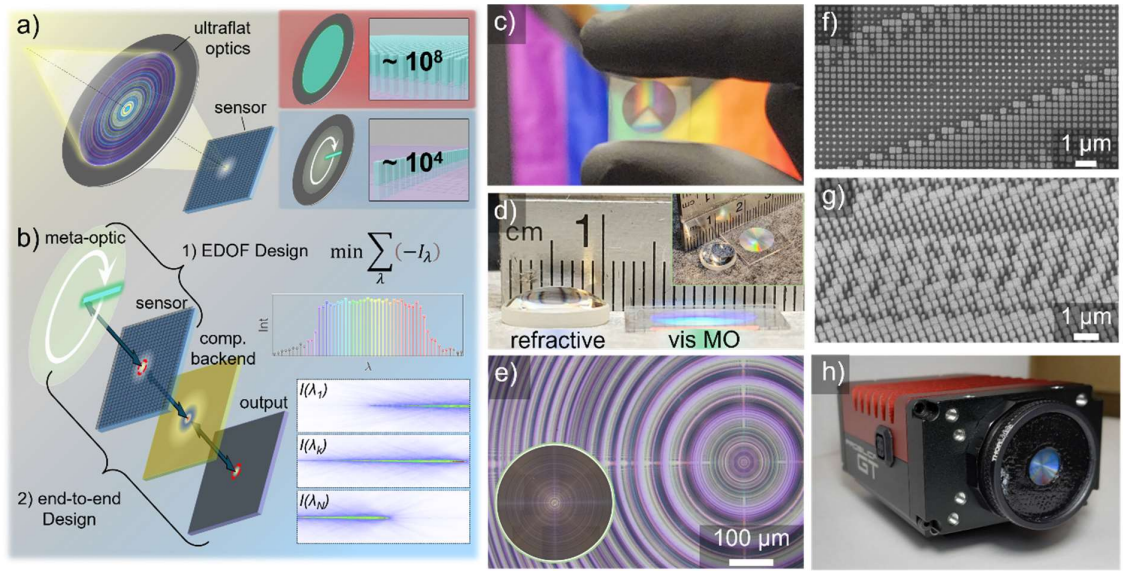

*Figure 1. Design and fabrication of the large aperture full-color meta-optic. a) Schematic description of the design challenge to achieve broadband focusing. By constraining the propagation to only the on-axis field, we compress the design space from a number of required scatterers of ~ $10^9$ to ~ $10^4$. b) The broadband capability is implemented in a two-step optimization process, where first a meta-optic is designed to exhibit maximal focal intensity for a dense sampling of wavelengths. Here, solely the intensity of the PSF for a broad spectral range at the center is maximized. This yields field profiles as shown by the intensity plots, along the optical axis. In the second step, we use this phase profile as an initialization to an end-to-end computational imaging optimization. c) Picture of the 1cm aperture meta-optic held*

*towards a vivid test scene, illustrating the direct lensing behavior. d) Side by side comparison of the meta-optic with a refractive lens of the same aperture and f-number. e) Optical microscope image of the center of the meta-optic. The inset shows a stitched image of the entire 1 cm aperture. f) and g) show scanning electron microscope (SEM) images of the scatterer in a top view and oblique view of 30°. h) The 1 cm aperture broadband meta-optic integrated with a camera using a 3D printed holder.*

Inverse design methods can yield functionalities that are often inaccessible through analytical solutions. However, the outcome is extremely reliant on the implemented figure of merit (FoM) and simulation method. As such many optimization problems do not ensure reaching the global minimum, and a poorly selected FoM can lead to optics which have only limited performance metrics. To understand what aspects constitute the strong performance of the designed meta-optic, we characterized all tested designs, including the initial (hyperboloid metalens), intermediate design (EDOF design), a full end-to-end polychromatic, and end-to-end broadband lenses. Their performance metrics are assessed against a refractive plano-convex lens of the same aperture and f-number with the same sensor and same computational back end architecture. This allows us to draw a direct comparison between the meta-optic and a refractive lens for the first time, deriving the same experimental metrics. By gauging the meta-optic against the refractive lens, we can accurately describe the potential of the large aperture broadband meta-optics.

First, we measured the PSF directly on sensor for a broad spectral range and angle of incidences (aoi). Light from a tunable source was collimated and focused by the meta-optic and directly captured at the design focal length, which remained fixed throughout the characterization (Figure 2a).

All measurements, including PSFs of other meta-optics are summarized in the Supplementary Information. Overall, we observe that throughout the experimentally accessible spectral range (480 nm – 680 nm) and for low aoi (0°, 5°), the refractive lens maintains a narrow PSF, which is strongly aberrated for larger angles (10°, 15°) (Figure 2b). In comparison, the fully optimized broadband end-to-end design exhibits a slightly more extended PSF at lower aoi, while maintaining a smaller PSF towards larger aoi (Figure 2c). This arises from the EDOF-like behavior of a broadband design, as further detailed in the Supplementary Information. A further relevant aspect of the broadband meta-optic design is the uniform PSF across the entire wavelength range. By deriving the modulation transfer function (MTF) we obtain insight into the variance of the PSF. Specifically, by extracting a threshold value, when the MTF contrast drops below a certain value (1%) we can identify the operating range for which a specific meta-optic is suitable and can yield a performance enhancement in imaging quality after computation. We observe that the end-to-end designed broadband meta-optic is well balanced with respect to the spectral range and aoi, and for broadband imaging it is closest to the refractive optics compared to other meta-optics tested. While other designs can achieve slightly higher quality at specific wavelengths or narrow bands, the broadband design achieves the highest performance overall over the entire spectral range.

To validate this claim, we directly measured the line pair contrast for the visible spectral range from 480 nm to 680 nm, in steps of 2 nm for all lenses, schematically illustrated in Figure 2d. For this measurement, light of varying wavelength passes through a diffusing medium and objective to ensure spatial and directionally uniform illumination of a standard resolution target (USAF 1951), which was placed in front of the optical element. We captured elements of groups 7, 6, 5 and 4 to obtain line pair contrasts across a large spatial frequency range from ~ 10 lp/mm up to ~ 150 lp/mm. Figure 2e, shows an example of a captured image with the broadband meta-optic at 580 nm. We note that the non-uniform appearance of the image likely occurred due to speckle pattern formation from the fiber coupling. Importantly, this measurement gives direct insight on the image performance enhancement once the computational backend is applied to the captured images (Fig 2f). As shown for a single wavelength (580 nm) in Figure 2g, the line pair contrast is increased by a factor of ~ 6 after computation and reaches values on par with the refractive lens. In comparison the refractive lens, or hyperboloid lens show almost no increase after computation. This directly demonstrates how different types of optics leverage the advantage provided by the computational reconstruction. The line pair contrast over the entire spectral range for different optics is plotted as colormaps in Figure 2h, summarizing the contrast values as function of wavelength, extracted from captured images, after deconvolution, and their respective difference, i.e., contrast enhancement. Notably, the hyperboloid metalens performs on par with the refractive lens only over a very narrow spectral band (~ 10 nm) and improvement after computation is only minimal. Opposite to that, we observe that the broadband meta-optic performs seemingly poorly for raw imaging over the entire visible range. The designed point spread function features a peak component and a tail which, without reconstruction, appears as aberrated component. Having preserved a peak component, after deconvolution, the design accounts the largest improvement with a contrast of about ~ 50% across the broad band range for line-pair contrast of ~ 10 – 50 lp/mm, on par performance with the refractive lens. The design effectively turns a deconvolution problem into a contrast enhancement problem. This finding validates the applicability for broadband, full-color imaging in consumer electronics, such as laptops, surveillance cameras, or smartphones, as the image information for these scenarios typically is in this contrast range, due to a sparser nature of image features. We note that some variations in plots appear due to varying position of the target elements in the respective field of views.

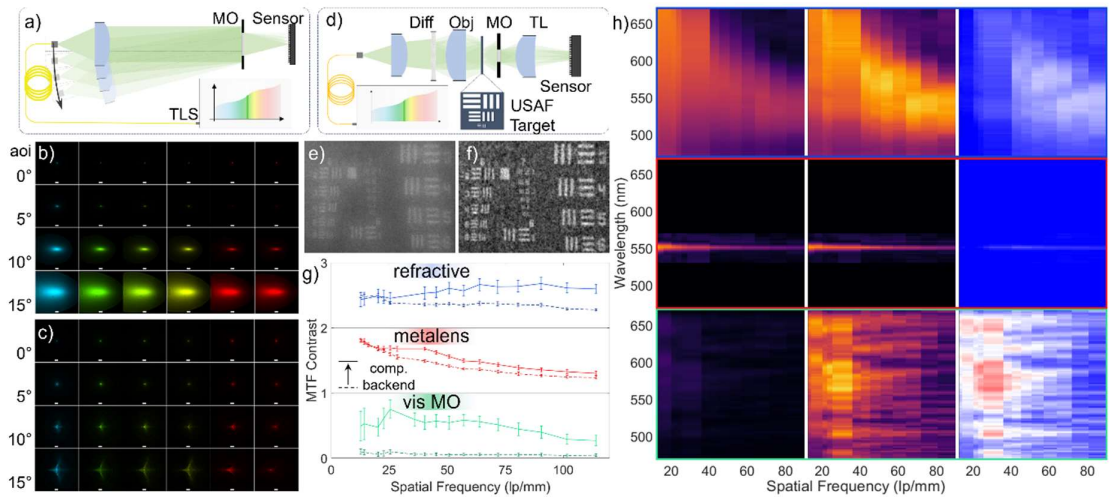

*Figure 2. Measurement of meta-optic performance. a) Schematic of the optical setup to measure the point spread function (PSF) for various optical elements under different angles of incidence (aoi). Comparison of measured PSFs for a plano-convex lens (b) and the broadband end-to-end meta-optic (c) for aoi of 0°, 5°,10°, and 15°. The images show the PSFs for wavelengths at 480 nm, 530 nm, 550 nm, 570 nm, 650 nm, 680 nm. Scale bars in the images correspond to 100 µm. d) Schematic of the optical setup to measure the contrast values for various optical elements for different wavelengths. (e) Raw image of the USAF 1951 target with elements of groups 5, 6, and 7 captured with the broadband end-to-end designed meta-optic for a wavelength of 580 nm. f) The same image after computational reconstruction is applied to the image. From images e) and f) the line pair contrasts are extracted. g) Examples of enhancement in MTF for line pair contrast before (dashed lines) and after (full lines) deconvolution for the refractive (top), hyperboloid (middle), and end-to-end design (bottom). h) Comparison of the obtained contrast values for line pairs are shown before (left column), and after deconvolution (middle column), as well as the relative difference (right column). From top down the plots correspond to the refractive lens, the hyperboloid metalens, and the end-to-end broadband meta-optic.*

To understand how these metrics translate to broadband imaging, we first captured images in a controlled environment, where an OLED screen was used to project a series of scenes in front of the camera (details in Method). Figure 3 summarizes a comparison of images after image reconstruction using a refractive lens (a), hyperboloid (b), and the broadband end-to-end meta-optic (c), respectively. Raw captures and images after specific operations are further discussed in Supplementary Information. To first assess the method without any learned priors, and recover the image purely based on a physical forward model, we employ here a conventional Wiener deconvolution and denoising(42) for captures. These were the same for the different optics, while specific parameters were adjusted to yield high quality imaging.

Under these conditions, the broadband meta-optic (c) provides overall imaging quality on-par with a refractive lens (a), with image details and specific color contrasts accurately captured. This broadband imaging capability is facilitated by the widely balanced MTF of the meta-optic. Further details, as shown in Figure (e) and (f) for the refractive lens and the broadband meta-optic, respectively, highlight the specific shortcomings of each system. Although the meta-optic can capture all colors and mid to large image features, fine structures such as the individual petals of the rose appear more blurred, compared to those of the refractive lens. In contrast, towards large field angles the meta-optic provides clearer image features as compared to those of the refractive lens. This enhanced performance at larger field angles is a direct consequence of the more confined PSF of the meta-optic compared to the refractive lens, as shown in Figure 2.

While the broadband meta-optic captures full-color images, the hyperboloid meta-optic is not suitable for generalized broadband imaging as only some color information is captured, which can be particularly well seen in images taken of the red tulips (top row). Although recent works have investigated neural back ends to reconstruct full-color images,(38, 39) we find that , with results collated in the Supplementary Information, for a large variety of scenes, the hyperboloid is not suitable, making this approach strongly scenario dependent. We again emphasize that the computational method used here is deliberately based on Wiener deconvolution and is therefore fully agnostic to the imaging scene as it only relies on the beforehand measured PSF. We also note that images captured with narrowband hyperboloids and neural backends in other works are largely sparse and have less color variation, which simplifies the reconstructions (further discussed in the Supplementary Information).

For a quantitative assessment of the broadband meta-optic, we measured the structural similarity index metric (SSIM) for a larger image set. By comparing the same images among the different lenses, using the same analysis pipeline, and using images with different scene features containing strong and weak contrasts, color, and brightness variations, we ensure a fair comparison of the respective imaging capabilities. The specific details of ensuring a high degree of overlapping features are described in the Supplementary Information. Averaged SSIMs for the different optics are shown in Figure 3(f), with values of ~ 0.65 for the refractive lens for RGB imaging, 0.6 for the broad band meta-optic, while the hyperboloid meta-optic only yields a value of ~ 0.55. For black and white imaging (after conversion to grayscale) the hyperboloid metalens in fact increases stronger in this metric. This shows that hyperboloid metalenses can be on par with other optics for black and white imaging, but for color information imaging over broadband are unsurprisingly ill suited.

While the OLED imaging experiment enables a quantitative assessment, cameras are commonly used under ambient illumination with light being reflected from a scene. This scenario drastically deviates from OLED imaging, where the image of a self-luminescent (pixeled) scene is captured. Therefore, to demonstrate the capability of broadband imaging in the presence of reflections, without narrow bandpass filters, we captured scenes indoor and outdoor. One such scene is shown in Figure 3(g), prepared to contain strong color and detail variations. Images after computation are shown in (h) and (i) for the meta-optic and the refractive lens respectively, with specific image details zoomed in. We find that the broadband MO captures the visible spectrum well, with similar detail close to the performance of the refractive lens. Whilst the image quality is on-par for small field angles, we observe that towards larger field angles, the meta-optic in fact outperforms the refractive lens, as image details are still recovered in the capture. This can be particularly well observed in the details, shown in Figures 3(h) and (i). Furthermore, we demonstrate video-rate, broadband imaging, suitable for video captures (full videos can be found in the Supplementary Information), shown in Figures 3(j)- 3(l). In the first video (room light, 5 MPix, captured with an exposure time of 90 ms, and frame rate of ~ 11 fps), a person is shown jumping in front of the same scene as Figure 3(g), with frames recording movements of the person, mid-jump and with recognizable features on their trousers. In the second video (room light, 5 MPix, captured with an exposure time of 5 ms and additional gain, and frame rate of ~ 19 fps), a person is shown flipping a coin, whereas the momentary position of the coin mid-air is captured. In the third video (daylight, 5 MPix, captured with an exposure time of 5 ms, and frame rate of ~ 19 fps) we demonstrate bouncing of a colorful ball, even capturing momentary deformations of the ball as it is thrown.

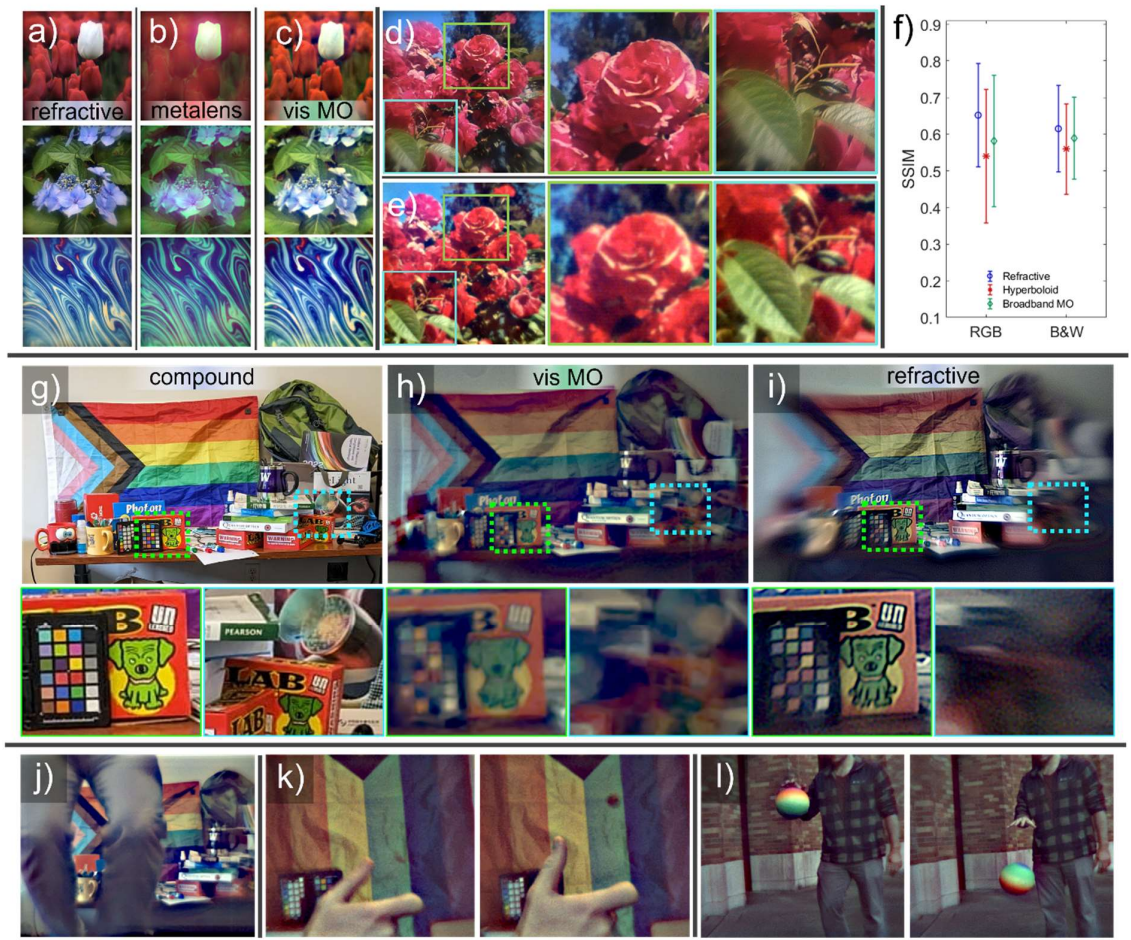

*Figure 3. Meta-optic imaging and comparison to refractive optics. In the first row, a set of images was projected on an OLED screen, and subsequently captured with various optical elements, and then reconstructed. To compare the design in the absence of learned priors, we employ here Wiener filtering as a physics-informed inverse filtering. Columns from left to right correspond to the computationally reconstructed images, after capturing with (a) refractive lens, (b) hyperboloid metalens, and (c) end-to-end designed meta-optic. Direct image comparison with details magnified for a refractive lens (d) and the end-to-end meta-optic (e). f) Averaged SSIM values for RGB images, as well as for a black and white (B&W). g) Scene prepared for comparison for real-life-scenario broadband imaging (no display capture). Images captured of the scene are shown in (h) and (i) using the broadband meta-optic and a refractive lens, respectively. Parts of the scene at the center (green outline) and off-axis (blue*

*outline) are magnified for better comparison of various details. j) Single frames from a series capturing a person jumping in front of the scene shown in (g). k) Frames showing a coin flip. l) Frames of a person bouncing a colorful ball, captured outside under daylight conditions.*

While the broadband meta-optic captures colors and scene details, it lacks the image quality achievable with state-of-the-art digital systems, such as smartphone cameras. Thus, to fully leverage the benefit of a computational backend and towards parity, we further augment the meta-optic with a learned reconstruction method. For advanced and spatially varying aberration correction, and better noise reduction, we designed a probabilistic diffusion-based neural network. To ensure the viability of the specific architecture for imaging under arbitrary conditions (ambient conditions and scene independence), we evaluate the method with a co-axial paired image capture system, schematically shown in Figure 4 (a). A beam splitter redirects 30% of light to a compound lens camera, while 70% of light is transmitted to a meta-optic camera. Both imaging systems share the same field of view, which ensures that reconstruction algorithms can be trained on arbitrary scenes. Details on reconstruction architecture and training are given in the Supplementary Information and Methods.

In Figure 4 (b), we directly compare ground truth captures (obtained via the compound lens imager), images reconstructed with a physics based inverse filter, and the learned backend. We note that the presented images were not part of the training set but were captured for verification of the approach. The learned probabilistic diffusion model outputs images which are significantly better than compared with the physics based inverse filter. Particularly, the images yield lower haze, better color accuracy, produce more vivid hues, better noise reduction, account for spatial variance, and the non-uniform sensor response. While parts of these factors could be improved as well in the physics-based backend, the probabilistic diffusion based reconstruction method directly takes these factors into account. Specifically, these factors can be particularly well seen in detail presented in zoomed-in sections in Figure 4(c). More images and details can be found in the Supplementary Information. As such, the broad-band meta-optic with the learned backend produces images, that are almost on-par with compound lens camera imagers.

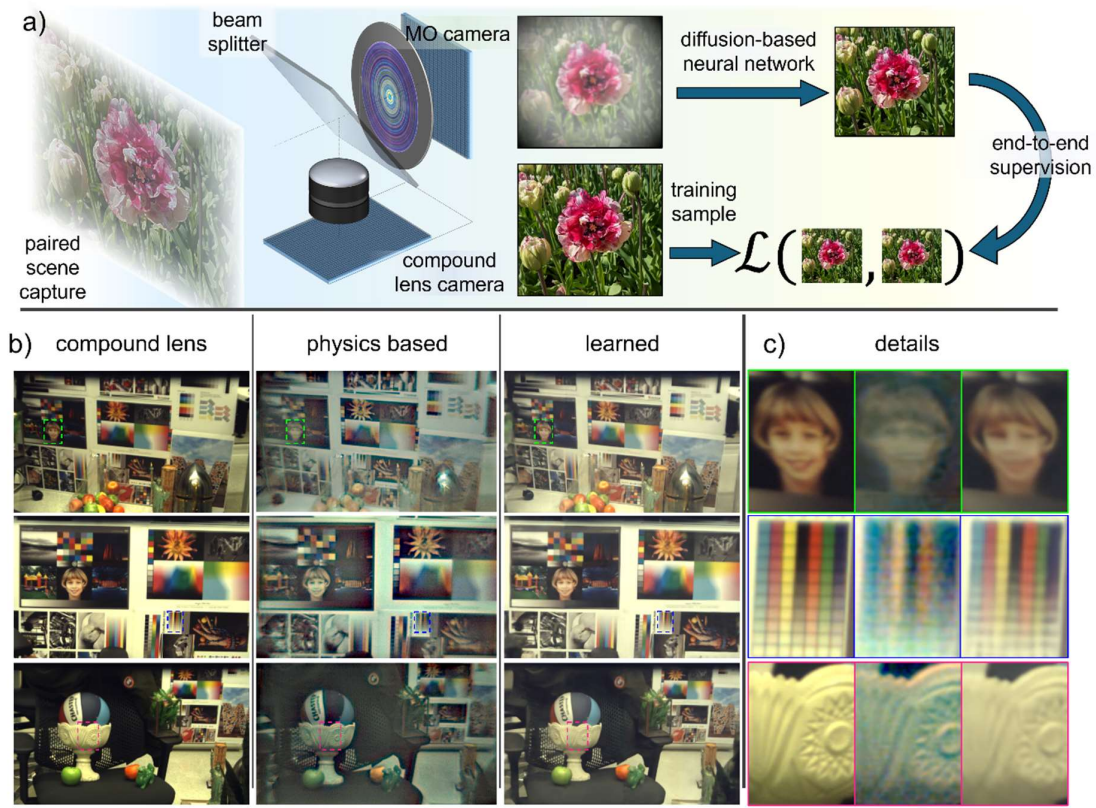

*Figure 4. Broadband imaging with a learned back end. a) Training a neural backend on images from real-life scenes. The same image scenes are captured with a meta-optical camera and a compound lens camera. A diffusion neural network is then trained and used to recover image quality. b) Comparison of images taken with a compound refractive lens, the broadband meta-optics and a physics based inverse filter, and with a learned backend. c) Zoomed in details of the highlighted boxes in the images are directly compared next to each other.*

**Discussion**

We demonstrate visible broadband, video-rate imaging with a single, polarization insensitive, 1 cm aperture, f/2, flat meta-optic. By designing a meta-surface in an end-to-end design fashion, we account for the inherent limitations of the flat diffractive optics in tandem with a computational reconstruction to ensure optimized performance within the provided constraints. This allows us to conclusively show how intrinsic chromatic limitations are overcome, and high-quality images are achievable with only a micron-thick diffractive layer. Through comparison with other designs, including the hyperboloid metalens, an EDOF design, and a polychromatic meta-optic, we validate the respective functionalities and how visible broadband imaging capability arises. Specifically, the formation of an extended depth of focus ensures all wavelengths within a defined spectral range are focused in a common plane. While this extends the PSF laterally, the final image quality is recovered through a deconvolution step. This is possible due to the well balanced PSF across the entire considered spectral range. We provide a full evaluation of the optical performance, and a head-to-head comparison with a refractive element to give a direct assessment of how well these optics compare with a refractive lens on the same sensor. Doing so, allows us to assess an upper limit on what the current state of the art is for an optical element of that specific size. Moreover, avoiding comparison with other meta-optics of smaller diameters circumvents the pitfall of scaled aberrations. As aberrations scale with the size of an optic, a larger optic will always perform more poorly in terms of its resolving power.(*40*, *41*) Beyond that, our system level comparison takes into account the limitations of the sensor itself, including the sensor MTF and angle of incidence dependence.

In short, we find on-par imaging performance with a single refractive lens, where we achieve an average MTF contrast after computation of ~ 50% in a range of 0 - 70 lp/mm, in comparison with an MTF contrast of ~ 60%, for a refractive lens in the same range (after computation). Moreover, we demonstrate improved imaging capabilities using a meta-optic at field angles of

10° and larger. The designed meta-optic is suitable for ambient imaging and video rate captures with short exposure times in the ms range, which we directly exhibit by capturing the motion of scenes with rapidly moving features. We ultimately achieve on par performance with compound imagers and demonstrate an approach to implement a learned reconstruction method using probabilistic diffusion models. Overall, this work demonstrates and elucidates in essence how broad band imaging in the visible is achievable with an ultra-flat optical element and large aperture (1 cm), and thus challenges the well-trenched belief in the community that broadband imaging is impossible using a large-aperture meta-optics.

**Methods**

**Meta-Optic Design**

Scatterers were designed used the S4 package for rigorous coupled wave analysis. (*43*)

**EDOF Design Approach**

In order to scale up to larger apertures we reduced the memory requirements by only simulating the center of the field of view. In doing so, we could employ rotationally symmetric image formation models that reduce computationally expensive 2D wave propagation into 1D systems. Coupled together with the RCWA approximation used from our previous work (*34*) it allowed for fast design of metalenses up to 5 cm in aperture diameter. In terms of design time, the optimization for a single 1 cm aperture metalens takes approximately one hour.

Another difference from our previous work is the sampling of wavelengths. We found that only 3 wavelengths (462nm, 511nm, 606nm) and using a polynomial basis for the metasurface design did not facilitate sufficient chromatic correction for larger aperture metasurfaces over the broad band. To remedy this, we instead sampled 1000 wavelengths within the 450 nm to

650 nm range and we used a per-pixel basis for the metasurface design. The per-pixel basis allow each nano-antenna along one radius of the metasurface to vary independently of the other nano-antennas without constraints. The figure of merit (FOM) used is maximization of focusing energy at the sensor plane for all wavelengths.

**End-to-End Approach**

We use the normal incidence PSF generated by the rotationally symmetric image formation model within the overall end-to-end design pipeline.(*34*) Specifically, we jointly design the metasurface together with the deconvolution network. The end-to-end optimization begins with the finetuned PSF obtained with the EDOF approach. We run the optimization process with joint optimization between the network and the metasurface for 1000 iterations.

**Fabrication**

A ~800 nm thick SiN film was deposited on a 300 μm thick quartz wafer using plasma enhanced chemical vapor deposition (PECVD) in a SPTS PECVD chamber. The wafer was then diced into 1.5 cm squares. After a brief cleaning (in Acetone and IPA) and barrel etch step ($O_2$, 100 W, 15s), a positive resist (ZEP 520 A) was spun onto the sample (4k rpm, thickness of ~ 400 nm), followed by baking at 180 °C for 3 min on a hot plate. A conductive polymer layer (DisCharge H2O) was subsequently spun on top at 4k rpm. The resist layer was then patterned using a 100 keV electron beam (JEOL JBX6300FS) at a dose of ~ 300 μC $cm^{-2}$. The total write time was about 4 1/2 hours. After EBL, the conductive polymer layer was removed in a short IPA bath and subsequently the resist was developed at room temperature in Amyl Acetate for 2 min. Subsequently, the sample was descummed in a short barrel etch step (100 W, 15s). Then, using electron beam evaporation, a layer of ~ 75 nm AlOx was deposited. The mask was then liftoff overnight in an NMP bath at ~ 100 C on a hot plate. Subsequently, the SiN layer was etched using a mixture of C4F8/SF6 in an inductively coupled reactive ion etcher

(Oxford PlasmaLab System 100). After etching the chip was subsequently integrated in a 3D printed holder and mounted in front of the sensor. For SEM imaging a thin conductive Au/Pd layer was deposited.

**Measurements**

The refractive lens used for comparison was purchase from Edmund Optics, specifically an uncoated Plano Convex lens with 1 cm diameter, 2 cm focal length.

To measure the PSFs of the optics at various wavelengths we used a custom built semi-automated setup, as schematically depicted in Figure 2a. Light of a broad band continuum source (NKT Photonics SUPERK Fianium FIR-20) is guided through an AOTF (NKT Photonics, SuperK Select) with selectable filter range, then coupled into one end of a single mode fiber (Thorlabs P1-630Y-FC-2). The other fiber end was mounted at a distance of ~ 50 cm away from the optic and placed on a custom built swing arm, which allows to change the angle of incidence onto the optics under test, while maintaining the same distance. The PSF was then directly captured on the sensor (Allied Vision ProSilica GT 1930, with a pixel size of ~ 5.6 µm), while maintaining the same distance from optic to sensor throughout the experiment. The wavelength and captures were automated using a custom script, which ensured that the respective PSF captures were not overexposed.

For MTF measurements (schematic, Figure 2d) the light of the broadband continuum source was coupled into a Multimode fiber (Thorlabs, 100 µm core size), and subsequently guided through a set of diffusers, and an objective (Mitutoyo, 10X, 0.28 NA) which illuminated a USAF 1951 Target (Thorlabs). The target was captured using the optics and a tube lens (Thorlabs TTL 180-A) was used to form an image on the sensor (Allied Vision ProSilica GT 1930, with a pixel size of ~ 5.6 µm). The line pair contrast was then calculated using an automated script as discussed in the Supplementary Information.

For image captures of a display, as shown in Figure 3 (a) - (e), a monitor was placed ~ 50 cm in front of the optic under test and a series of images was displayed with an image height/width of ~ 16 cm, corresponding to a horizontal FoV of 20°. Images were captured using an automated script, and a Allied Vision ProSilica GT 1930. Some images were taken from the INRA Holiday set.(*44*)

For image captures under ambient conditions we prepared a colorful scene as shown in Figure 3(g), image captured with a Google Pixel 6A. Other images were captured with a meta-optic or refractive lens and a ZWO ASI183MC Pro camera. We note that for all captures a broadband filter (400 nm - 750 nm) was added in the camera to eliminate undesired IR illumination.

For paired image capture, we built a portable custom capture system, with two-coaxially aligned cameras, one aligned with the broadband meta-optics, one with a compound lens. Both cameras shared the same scene using a 70/30 beam splitter, whereas 70 % of light is transmitted to the MO camera, while 30 % of light is reflected to the compound lens camera. Both cameras used the same sensor (Allied Vision GT 1930c). We first aligned the imaging system on checkerboard patterns, which ensured that the same image scene would be captured and enable the accurate learning and testing of this imaging system. More details on the paired image capture method can also be found in Ref (*36*).

**Acknowledgements**

Part of this work was conducted at the Washington Nanofabrication Facility/Molecular Analysis Facility, a National Nanotechnology Coordinated Infrastructure (NNCI) site at the

University of Washington, with partial support from the National Science Foundation via awards NNCI-1542101, NNCI-2025489, and DARPA W31P4Q21C0043. Felix Heide was supported by a NSF CAREER Award (2047359), a Packard Foundation Fellowship, a Sloan Research Fellowship, a Sony Young Faculty Award, and an Amazon Science Research Award.

**References**

1. J. N. Mait, G. W. Euliss, R. A. Athale, Computational imaging. *Adv. Opt. Photon., AOP* **10**, 409–483 (2018).

2. M. Delbracio, D. Kelly, M. S. Brown, P. Milanfar, Mobile Computational Photography: A Tour. *Annual Review of Vision Science* **7**, 571–604 (2021).

3. J. Suo, W. Zhang, J. Gong, X. Yuan, D. J. Brady, Q. Dai, Computational Imaging and Artificial Intelligence: The Next Revolution of Mobile Vision. arXiv arXiv:2109.08880 [Preprint] (2021). https://doi.org/10.48550/arXiv.2109.08880.

4. D. N. Neshev, A. E. Miroshnichenko, Enabling smart vision with metasurfaces. *Nat. Photon.* **17**, 26–35 (2023).

5. J. Hu, S. Bandyopadhyay, Y. Liu, L. Shao, A Review on Metasurface: From Principle to Smart Metadevices. *Frontiers in Physics* **8** (2021).

6. M. Pan, Y. Fu, M. Zheng, H. Chen, Y. Zang, H. Duan, Q. Li, M. Qiu, Y. Hu, Dielectric metalens for miniaturized imaging systems: progress and challenges. *Light Sci Appl* **11**, 195 (2022).

7. J. An, K. Won, Y. Kim, J.-Y. Hong, H. Kim, Y. Kim, H. Song, C. Choi, Y. Kim, J. Seo, A. Morozov, H. Park, S. Hong, S. Hwang, K. Kim, H.-S. Lee, Slim-panel holographic video display. *Nat Commun* **11**, 5568 (2020).

8. X. Fang, H. Ren, M. Gu, Orbital angular momentum holography for high-security encryption. *Nat. Photonics* **14**, 102–108 (2020).

9. G.-Y. Lee, J.-Y. Hong, S. Hwang, S. Moon, H. Kang, S. Jeon, H. Kim, J.-H. Jeong, B. Lee, Metasurface eyepiece for augmented reality. *Nat Commun* **9**, 4562 (2018).

10. S. Colburn, A. Majumdar, Metasurface Generation of Paired Accelerating and Rotating Optical Beams for Passive Ranging and Scene Reconstruction. *ACS Photonics* **7**, 1529–1536 (2020).

11. Q. Guo, Z. Shi, Y.-W. Huang, E. Alexander, C.-W. Qiu, F. Capasso, T. Zickler, Compact single-shot metalens depth sensors inspired by eyes of jumping spiders. *Proceedings of the National Academy of Sciences* **116**, 22959–22965 (2019).

12. Z. Lin, R. Pestourie, C. Roques-Carmes, Z. Li, F. Capasso, M. Soljačić, M. Soljačić, S. G. Johnson, End-to-end metasurface inverse design for single-shot multi-channel imaging. *Opt. Express, OE* **30**, 28358–28370 (2022).

13. N. A. Rubin, G. D'Aversa, P. Chevalier, Z. Shi, W. T. Chen, F. Capasso, Matrix Fourier optics enables a compact full-Stokes polarization camera. *Science* **365**, eaax1839 (2019).

14. Z. Shen, F. Zhao, C. Jin, S. Wang, L. Cao, Y. Yang, Monocular metasurface camera for passive single-shot 4D imaging. *Nat Commun* **14**, 1035 (2023).

15. J. Qin, S. Jiang, Z. Wang, X. Cheng, B. Li, Y. Shi, D. P. Tsai, A. Q. Liu, W. Huang, W. Zhu, Metasurface Micro/Nano-Optical Sensors: Principles and Applications. *ACS Nano* **16**, 11598–11618 (2022).

16. Y. Intaravanne, X. Chen, Recent advances in optical metasurfaces for polarization detection and engineered polarization profiles. *Nanophotonics* **9**, 1003–1014 (2020).

17. M. Faraji-Dana, E. Arbabi, A. Arbabi, S. M. Kamali, H. Kwon, A. Faraon, Compact folded metasurface spectrometer. *Nat Commun* **9**, 4196 (2018).

18. J. E. Fröch, S. Colburn, A. Zhan, Z. Han, Z. Fang, A. Saxena, L. Huang, K. F. Böhringer, A. Majumdar, Dual Band Computational Infrared Spectroscopy via Large Aperture Meta-Optics. *ACS Photonics*, doi: 10.1021/acsphotonics.2c01017 (2022).

19. R. Wang, M. A. Ansari, H. Ahmed, Y. Li, W. Cai, Y. Liu, S. Li, J. Liu, L. Li, X. Chen, Compact multi-foci metalens spectrometer. *Light Sci Appl* **12**, 103 (2023).

20. Z. Yang, T. Albrow-Owen, W. Cai, T. Hasan, Miniaturization of optical spectrometers. *Science* **371**, eabe0722 (2021).

21. K. Wei, X. Li, J. Froech, P. Chakravarthula, J. Whitehead, E. Tseng, A. Majumdar, F. Heide, Spatially Varying Nanophotonic Neural Networks. arXiv arXiv:2308.03407 [Preprint] (2023). https://doi.org/10.48550/arXiv.2308.03407.

22. H. Zheng, Q. Liu, Y. Zhou, I. I. Kravchenko, Y. Huo, J. Valentine, Meta-optic accelerators for object classifiers. *Science Advances* **8**, eabo6410 (2022).

23. L. Huang, Q. A. A. Tanguy, J. E. Froch, S. Mukherjee, K. F. Bohringer, A. Majumdar, Photonic Advantage of Optical Encoders. arXiv arXiv:2305.01743 [Preprint] (2023). https://doi.org/10.48550/arXiv.2305.01743.

24. M. K. Chen, X. Liu, Y. Sun, D. P. Tsai, Artificial Intelligence in Meta-optics. *Chem. Rev.* **122**, 15356–15413 (2022).

25. E. Arbabi, A. Arbabi, S. M. Kamali, Y. Horie, A. Faraon, Multiwavelength polarization-insensitive lenses based on dielectric metasurfaces with meta-molecules. *Optica, OPTICA* **3**, 628–633 (2016).

26. L. Huang, S. Colburn, A. Zhan, A. Majumdar, Full-Color Metaoptical Imaging in Visible Light. *Advanced Photonics Research* **n/a**, 2100265.

27. F. Presutti, F. Monticone, Focusing on bandwidth: achromatic metalens limits. *Optica, OPTICA* **7**, 624–631 (2020).

28. J. Engelberg, U. Levy, Achromatic flat lens performance limits. *Optica, OPTICA* **8**, 834–845 (2021).

29. R. J. Lin, V.-C. Su, S. Wang, M. K. Chen, T. L. Chung, Y. H. Chen, H. Y. Kuo, J.-W. Chen, J. Chen, Y.-T. Huang, J.-H. Wang, C. H. Chu, P. C. Wu, T. Li, Z. Wang, S. Zhu, D. P. Tsai, Achromatic metalens array for full-colour light-field imaging. *Nat. Nanotechnol.* **14**, 227–231 (2019).

30. S. Wang, P. C. Wu, V.-C. Su, Y.-C. Lai, M.-K. Chen, H. Y. Kuo, B. H. Chen, Y. H. Chen, T.-T. Huang, J.-H. Wang, R.-M. Lin, C.-H. Kuan, T. Li, Z. Wang, S. Zhu, D. P. Tsai, A broadband achromatic metalens in the visible. *Nature Nanotech* **13**, 227–232 (2018).

31. W. T. Chen, A. Y. Zhu, V. Sanjeev, M. Khorasaninejad, Z. Shi, E. Lee, F. Capasso, A broadband achromatic metalens for focusing and imaging in the visible. *Nature Nanotech* **13**, 220–226 (2018).

32. W. T. Chen, A. Y. Zhu, F. Capasso, Flat optics with dispersion-engineered metasurfaces. *Nat Rev Mater* **5**, 604–620 (2020).

33. J. E. Fröch, L. Huang, Q. A. A. Tanguy, S. Colburn, A. Zhan, A. Ravagli, E. J. Seibel, K. F. Böhringer, A. Majumdar, Real time full-color imaging in a Meta-optical fiber endoscope. *eLight* **3**, 13 (2023).

34. E. Tseng, S. Colburn, J. Whitehead, L. Huang, S.-H. Baek, A. Majumdar, F. Heide, Neural nano-optics for high-quality thin lens imaging. *Nat Commun* **12**, 6493 (2021).

35. A. McClung, S. Samudrala, M. Torfeh, M. Mansouree, A. Arbabi, Snapshot spectral imaging with parallel metasystems. *Science Advances* **6**, eabc7646 (2020).

36. P. Chakravarthula, J. Sun, X. Li, C. Lei, G. Chou, M. Bijelic, J. Froesch, A. Majumdar, F. Heide, Thin On-Sensor Nanophotonic Array Cameras. *ACM Trans. Graph.* **42**, 249:1-249:18 (2023).

37. A. V. Baranikov, E. Khaidarov, E. Lassalle, D. Eschimese, J. Yeo, N. D. Loh, R. Paniagua-Dominguez, A. I. Kuznetsov, Large field-of-view and multi-color imaging with GaP quadratic metalenses. arXiv arXiv:2305.16676 [Preprint] (2023). https://doi.org/10.48550/arXiv.2305.16676.

38. R. Maman, E. Mualem, N. Mazurski, J. Engelberg, U. Levy, Achromatic imaging systems with flat lenses enabled by deep learning. arXiv arXiv:2308.12776 [Preprint] (2023). https://doi.org/10.48550/arXiv.2308.12776.

39. Y. Dong, B. Zheng, H. Li, H. Tang, Y. Huang, S. An, H. Zhang, High-fidelity achromatic metalens imaging via deep neural network. arXiv arXiv:2308.00211 [Preprint] (2023). https://doi.org/10.48550/arXiv.2308.00211.

40. A. W. Lohmann, Scaling laws for lens systems. *Appl. Opt., AO* **28**, 4996–4998 (1989).

41. D. J. Brady, N. Hagen, Multiscale lens design. *Opt. Express, OE* **17**, 10659–10674 (2009).

42. Y. Mäkinen, L. Azzari, A. Foi, Collaborative Filtering of Correlated Noise: Exact Transform-Domain Variance for Improved Shrinkage and Patch Matching. *IEEE Transactions on Image Processing* **29**, 8339–8354 (2020).

43. V. Liu, S. Fan, S4 : A free electromagnetic solver for layered periodic structures. *Computer Physics Communications* **183**, 2233–2244 (2012).

44. H. Jegou, M. Douze, C. Schmid, "Hamming Embedding and Weak Geometric Consistency for Large Scale Image Search" in *Computer Vision – ECCV 2008*, D. Forsyth, P. Torr, A. Zisserman, Eds. (Springer, Berlin, Heidelberg, 2008)*Lecture Notes in Computer Science*, pp. 304–317.